\documentclass{acm_proc_article-sp} 
\usepackage{placeins,subfigure,float,array,booktabs} 
\usepackage[font=small,labelfont=bf]{caption} 
\begin{document} 
\title{Randomised Relevance Model} 
%
%
%
%
%
\numberofauthors{3}  
\author{ 
\alignauthor Dominik Wurzer\\ 
\affaddr{School of Informatics}\\ 
\affaddr{University of Edinburgh}\\ 
\email{D.S.Wurzer@sms.ed.ac.uk} 
\alignauthor Miles Osborne\\ 
\affaddr{School of Informatics}\\ 
\affaddr{University of Edinburgh}\\ 
\email{miles@inf.ed.ac.uk} 
\alignauthor Victor Lavrenko\\ 
\affaddr{School of Informatics}\\ 
\affaddr{University of Edinburgh}\\ 
\email{vlavrenk@inf.ed.ac.uk} 
} 
\maketitle 
\begin{abstract} 
Relevance Models are well known retrieval models and capable of producing competitive results. However, because they use query expansion they can be very slow. We address this slowness by incorporating two variants of locality sensitive hashing (LSH) into the query expansion process. Results on two document collections suggest that we can obtain large reductions in the amount of work, with a small reduction in effectiveness. Our approach is shown to be additive when pruning query terms. 
\end{abstract} 
\section{Introduction} 
The Relevance Model (RM) is a retrieval model which produces competitive results, in part because it incorporates massive query expansion into the retrieval process ( Lavrenko \& Croft, 2001).  
The relevance-modelling approach to retrieval works in two steps: 
\\ 
{\bf 1.} Compute the conditional probability $P(w|Q)$ for observing a 
word $w$ as a possible extension of the query $Q$. A collection of 
these probabilities is called the "relevance model". 
\\ 
{\bf 2.} Rank every document by the Kullback-Leibler divergence 
between its language model and the relevance model computed in step 1. 
Unfortunately, Relevance Models are slow. Lavrenko \& Allan (2006) and Cartright et al. (2010) identified two factors as the computational bottleneck: the number of terms in the relevance model ($qs$), and the number of documents in the index ($nd$).  
The most common way to speed up RMs is to prune the number of expansion terms, which reduces the query size ($qs$) in the second retrieval step (Metzler et al., 2005; Lavrenko \& Allan, 2006). Carpineto \& Romano (2012) however criticised this threshold pruning technique because although it speeds up RMs, it can reduce effectiveness. Additionally pruning has no theoretical connection with information retrieval. Lavrenko \& Allan (2006) proposed to shift the workload from the runtime to the indexing time. They pre-compute the degree of relatedness between all documents to form a similarity index. This structure is looked up at runtime, after the initial retrieval step, to identify further relevant documents. Their algorithm reduces the runtime of RMs to the level of Language Models but requires vast computational effort at indexing time, which causes difficulties when it is applied to larger corpora (Carpineto \& Romano, 2012).  
The Randomised Relevance Model (RRM) increases the efficiency of Relevance Models by using Locality Sensitive Hashing to reduce the number of documents ranked in the final step. We investigate two approaches - a basic LSH scheme, which speeds up the final retrieval step and a novel approach based upon Multi Probe LSH (MPL). We demonstrate reductions in the amount of computation needed, with only minor reductions in effectiveness. Importantly our approach is additive with pruning. 
\section{Randomised Relevance Model} 
Relevance Models carry out multiple retrieval steps. In particular, the final step (which evaluates the expanded query) can cause a significant slowdown. An inverted index allows us to rank a collection of documents with an average-case complexity of $O(qs * nd * ds / vs)$ operations. Here $qs$ refers to the query size, $nd$ is the number of documents in our index, $ds$ is the average document size and $vs$ is the number of distinct terms in the index. Our approach, called Randomised Relevance Model (RRM), makes the final retrieval step efficient. We use Locality Sensitve Hashing (LSH) by Charikar (2002) to recover potentially relevant documents, which reduces the number of documents (\textit{nd}) each query is compared with. LSH is a randomized algorithm that identifies similar objects by assigning them identical hash-codes based on their position within the Vector Space. 
We use a standard LSH scheme (Charikar, 2002) to assign hash-codes to each document in the collection, placing them into buckets. At retrieval time, we compute the hash-code of the expanded query and match it against our collection, yielding a bucket of candidate documents. These documents are then used by the final retrieval step. Since (depending upon parameterisation) this set of candidate documents is typically small, we are able to rank them faster.  
Figure 1 a) visualizes this process using 3 hash functions ($h1$ - $h3$), identifying document $D1$ in bucket $B1$ as related to the query $Q$. For this set-up, LSH fails to retrieve document $D2$, although it is close to the query. As is standard, we make use of multiple hash-tables to reduce false negative errors by LSH (Petrovic et al., 2010). That is, at times two documents which should be in the same bucket  
are assigned different hash-codes, meaning that they do not collide (and hence are considered as not being near neighbours of each other). By repeating this process as shown in Figure 1 b), we reduce the chance of this happening. This drives down the error rate at the cost of more time and space --we need to build and visit more tables. 
\begin{figure*}[ht] 
\centering 
\subfigure[$1^{st}$ Hash-table of basic LSH using 3 bits]{ 
\epsfig{file=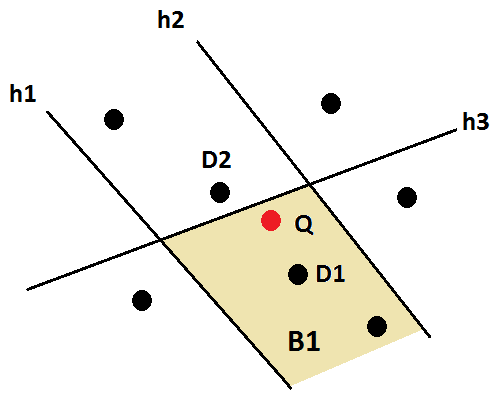, height=1.5in, width=1.5in} 
}%
\hspace{1.3mm} 
\subfigure[$2^{nd}$ Hash-table of basic LSH using 3 bits]{ 
\epsfig{file=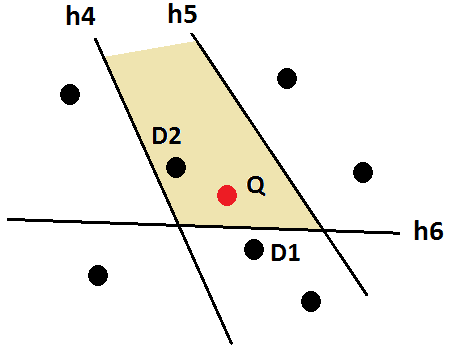, height=1.5in, width=1.5in} 
}%
\hspace{1.4mm} 
\subfigure[MPL using 1 hash-table, 3 bits and 1 probe]{ 
\epsfig{file=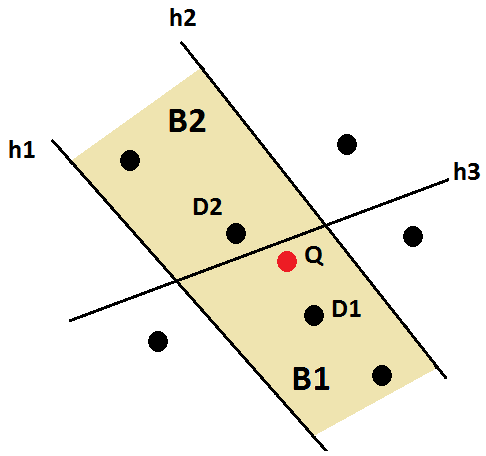, height=1.6in, width=1.6in} 
}%
\hspace{1.3mm} 
\subfigure[MP-RRM using 1 hash-table, 5 bits and 1 probe]{ 
\epsfig{file=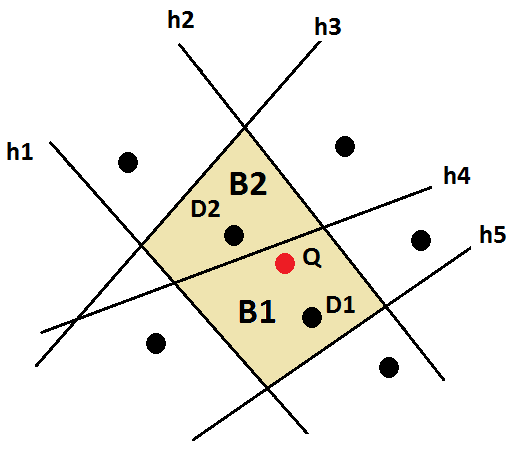, height=1.6in, width=1.6in} 
}%
\caption{Illustrating the principles of basic LSH, MPL and MP-RRM on the same data-set} 
\end{figure*} 
\section{Multi Probe Randomised \\Relevance Model} 
Multi Probe LSH (MPL) by Lv et al. (2007) reduces the amount of memory used by LSH. The algorithm makes efficient use of fewer hash-tables by probing additional "nearby" buckets within each table. The principle is illustrated by Figure 1 c), where MPL is able to retrieve document $D1$ and $D2$ by using only a single hash-table but probing two buckets ($B1$ and $B2$). The probing sequence is directed by the shortest hamming distance between the hash-code of the query and hash-codes corresponding with adjacent bucket. This acts as an approximation of whether the bucket is likely to contain documents that are close to the query. Every inspection of additional buckets increases the likelihood of identifying sought documents. As a consequence, MPL overcomes the need for inefficient additional hash-tables, without increasing the error rate. 
We adapt the MPL approach to Randomised Relevance Models but in contrast to the work of Lv, we do not use MPL to reduce the amount of memory to run LSH. Instead, we enhance the speed of RRM by further reducing the number of documents ranked. To do so, we configure the Multi Probe RRM (MP-RRM) using relatively many hash functions, as seen in Figure 1 d). This tends to produce buckets with fewer documents in them. At run time and when needed, by probing nearby buckets we effectively merge them. In the high dimensional Vector Space, every bucket with a hamming distance of 1 to the query bucket becomes a potential probing candidate and is considered as an adjacent bucket. In practice we tend to need only a small number of additional probing steps when reaching a desired effectiveness level. The total number of documents we consider using MP-LSH tends to be lower than when using a traditional LSH scheme, achieving the same effectiveness levels. As a result MP-RRM further reduces the amount of documents to be ranked, which increases its efficiency compared to plain RRM. The number of probing steps trades effectiveness against efficiency and is another parameter to be tuned. 
\section{Evaluation} 
\subsection{Data Set}  
We evaluate the RRM and MP-RRM on two different collections, the TDT1 (Wayne, 1998) and TREC-Robust (Voorhees, 2004). TDT1 contains 15,861 English news wire stories, 67,710 distinct terms and provides 25 official queries with exhaustive relevance judgements. TREC-ROBUST consists of 528,155 documents and 1,101,371 distinct terms. We make use of the 50 official queries and their relevance judgements, that were labelled as hard. These queries consist of the 50 hardest queries from previous TREC corpora and represent the characteristics of web search queries.  
\subsection{Evaluation Metrics} 
When evaluating the RRM and MP-RRM, we focus on a setting targeted at a web user. Therefore, we measure effectiveness by Precision at rank 5, as is standard for high precision scenarios. This is a reasonable metric, considering that 88\% of web users click on the 5 highest ranked documents (Enge et al., 2009). \\ 
The massively expanded query is the dominant factor on the runtime of Relevance Models. Commonly used efficiency metrics such as dot-products ignore the query length and are therefore not able to represent the actual efficiency of the algorithm. We evaluate efficiency using two different metrics: \emph{wall clock} 
and \emph{complexity}. Wall clock represents the actual running time of the algorithm in seconds, averaged over 5 runs. As the runtime is highly system-dependant, we also measure the average computational \emph{complexity} of our approach: the aggregate number of operations required to rank all documents with respect to a given query. This complexity figure is exactly the same as the $O(qs * nd * ds / vs)$ expression we provided in section 2. \\\\
\subsection{Experiments} 
Our baseline model implements the standard Relevance Model (RM), as described by Lavrenko \& Croft (2001). 
LSH allows RRM and MP-RRM to be parameterised in terms of a time/error trade-off. Long hash-codes produce lower complexity but RRM and MP-RRM suffer a drop in effectiveness. Shorter hash-codes ensure high effectiveness but reduced efficiency. In Table 1 we report the set-up in the algorithm label, which consists of the algorithm name, followed by the number of expansion terms, the number of bits per hash-code, the number of hash-tables and the number of probes. If not stated differently, the default setup for RRM and MP-RRM use 200 expansion terms and 18 hash-tables, which provide a reasonable balance between the required space, effectiveness and efficiency. \\ 
Configuring the LSH models requires selecting the number of hash functions and tables. This can be achieved using held-out data. Because we have relatively few queries we could not do this and so report results optimised on the test set. As shown by Figures 2 and 4, varying the number of hash functions and tables and probes produces the theoretically expected results: we can trade efficiency for effectiveness. We do not find for example that the RM is ever more efficient and effective than the RRM or MP-RRM. \\\\
\begin{figure}[ht] 
\centerline{\epsfig{file=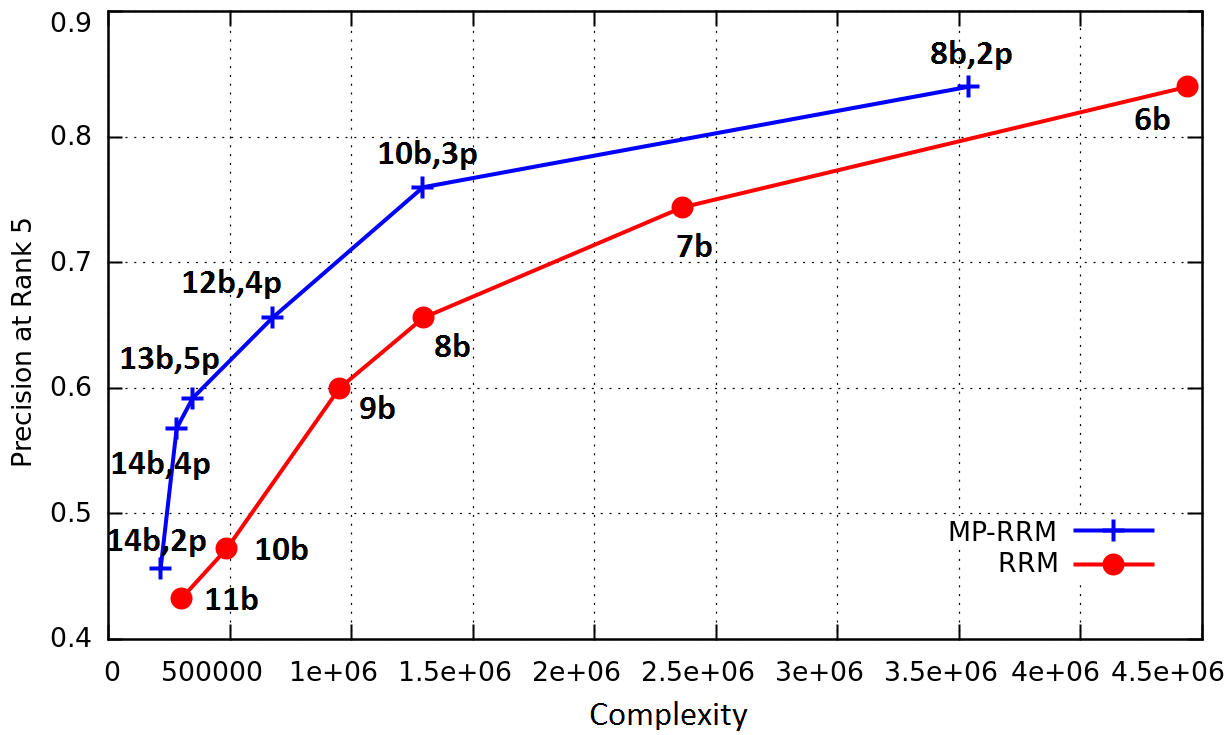, scale=0.28}} 
\vspace{-3mm} 
\caption{Effectiveness-Efficiency trade-off curve for TDT1; showing the impact of bit-length (b) and probes (p) on RRM and MP-RRM; both using 18 hash tables and 1000 expansion terms}  
\end{figure} 
\begin{figure}[ht] 
\centerline{\epsfig{file=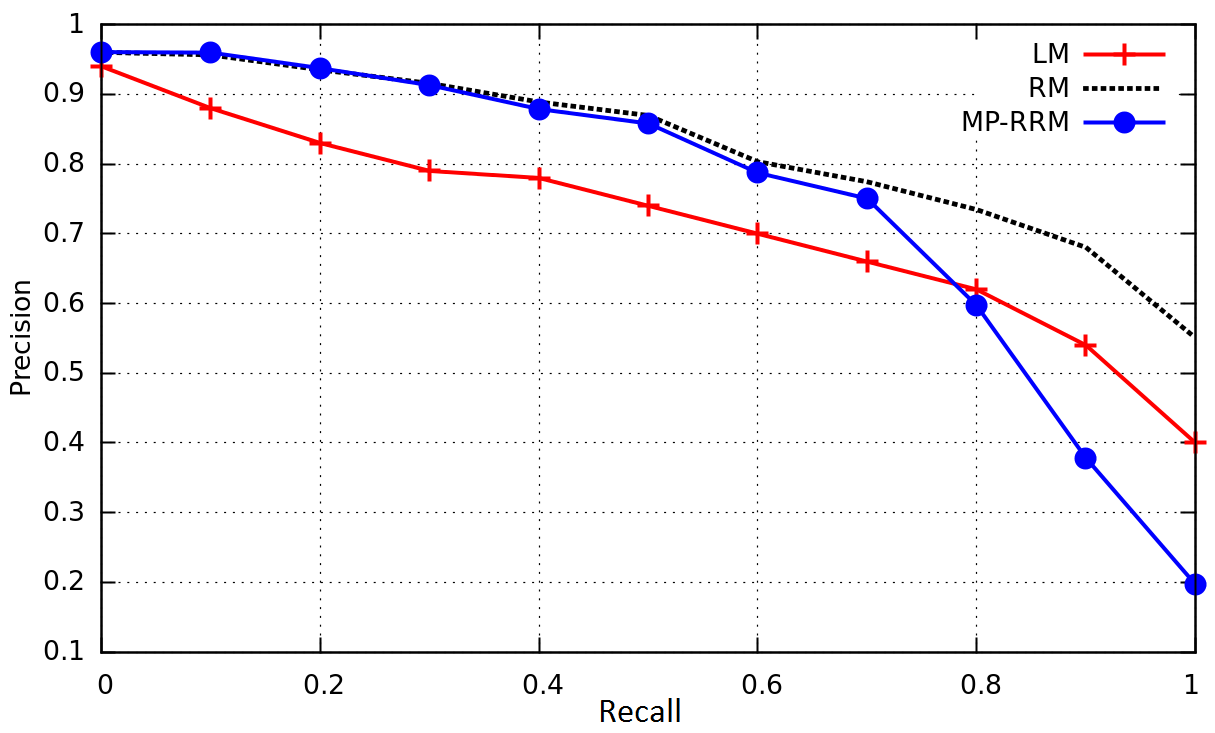, scale=0.28}}  
\vspace{-3mm} 
\caption{Interpolated Recall Precision Plot for MP-RRM using 100 terms, 8 bit, 18 hash-table, 3 probes and RM using 200 terms as well as a Language Model for TDT1}  
\end{figure} 
\textbf{RM vs. RRM vs. MP-RRM} \\ 
RRM shows an increase in efficiency over RM, but in particular on TDT1, its impact is limited. We find the low number of bits per hash-code to be the main cause of its performance. Figure 2 shows that RRM requires a low number of bits to reach adequate effectiveness. This is caused by LSH, which finds documents nearest to the query. RRM however, hashes an expanded RM query, which is not guaranteed to be close to all relevant documents. Hence, RRM is forced to widen its buckets to gather enough relevant documents. RRM is therefore compelled to use a few big buckets, which prevent it from reaching its full potential to speed up RMs. \\ 
The standard approach of retaining low error rates with small buckets is to increase the number of hash-tables. None-theless, constructing and visiting more hash-tables would increases the computational effort of RRMs. 
\\ 
In comparison with RRM, we configure MP-RRM to use more bits per hash-code. Every additional bit cuts the average number of documents per bucket in half. The final bucket is then built by combining the buckets closest to the query. MP-RRM lowers the complexity by 37.5\% on TDT1 and 42.24\% on TREC-ROBUST in comparison with RRM. The reduced computational effort allows MP-RRM to perform faster than RRM for every level of effectiveness, as shown in Figure 2. While our experiments evaluate effectiveness by Precision at Rank 5, Figure 3 shows that MP-RRM results in effectiveness comparable to RM for up to 60\% of recall. To conclude these experiments, basic RRM results in modest gain over RM. However, MP-RRM can offer considerable gains in efficiency for typical web search scenarios. 
\begin{table*}[!htbp]  
\small  
\centering  
\begin{tabular}{|c|c|c|c|c|c|c|c|c|}  
\multicolumn{7}{c}{\textbf{TDT1}}\\  
\hline  
Algorithm & P@5 & Difference & Complexity & Difference & Seconds & Difference \\  
& & of P@5 in \%& & of Complexity in \% & & of Seconds in \% \\  
\hline  
RM-baseline (200) & 0.864 & - & 14,315,788 & - & 35 & - \\  
RRM (200,6,18) & 0.848 & -1.85 & 7,418,506 & -48.18 & 20 & -42.86 \\  
MP-RRM (200,9,18,4) & 0.848 & -1.85 & 4,636,291 & -67.61 & 13 & -62.85 \\  
pruned RM (16) & 0.822 & -4.86 & 2,001,548 & -86.02 & 5 & -85.71 \\  
pruned MP-RRM (100,8,18,3) & 0.864 & 0 & 1,992,440 & -86.08 & 6 & -82.86\\  
\toprule  
\multicolumn{7}{c}{\textbf{TREC-ROBUST}}\\  
\hline  
RM-baseline (1000) & 0.324 & - & 589,340,657 & - & 1,712 & - \\  
RRM (1000,7,18) & 0.32 & -1.24 & 198,579,495 & -66.31 & 695 & -59.4 \\  
MP-RRM (1000,10,18,4) & 0.316 & -2.47 & 114,685,613 & -80.54 & 478 & -72.08 \\  
pruned RM (100) & 0.292 & -9.88 & 112,354,284 & -80.94 & 451 & -73.66 \\  
pruned MP-RRM (800,7,18,4) & 0.324 & 0 & 111,642,265 & -81.06 & 471 & -72.49\\  
\hline  
\end{tabular}  
\caption{Comparison of RRM, Multi Probe RRM and RM on the TDT1 and TREC-ROBUST data set;  
Algorithm labels consists of the algorithm name, followed by the number of expansion terms, bits per hash-code, hash-tables and probes}  
\end{table*} 
\\ \\ \textbf{Alternative Speed-up} \\ 
We compare the MP-RRM against the pruning of the number of expansion terms, which is the most commonly applied technique to speed up RMs. The result of pruning the expansion terms is a reduction in distinct query terms $(\textit{qs})$ in expression $O(qs * nd * ds / vs)$, which we provided in section 2. This method is also applicable to RRM and MP-RRM, whereas here, the number of distinct query terms $(\textit{qs})$ and documents in the index $(\textit{nd})$ is reduced. When pruning the number of expansion terms, we generally observe an increase in efficiency, at the cost of effectiveness, as described by Lavrenko \& Allan (2006) and Carpineto \& Romano (2012). However, each algorithm reveals a different optimum number of expansion terms dependent on the corpora. Figure 4 shows that the MP-RRM is able to compensate for a reduction in efficiency caused by fewer expansion terms, to a certain extent through a different configuration. Slightly wider buckets allow the retrieval algorithm to reach an adequate level of effectiveness with fewer expansion terms, which causes additionally increased efficiency in comparison with RM. Our experiments reveal that pruned MP-RRM is able to reduce the amount of operations of RM by 86.08\% on TDT1 and 81.06\% on TREC-ROBUST, while retaining the same level of effectiveness. Comparable speed-up is also achievable through conventional expansion term pruning on RM. However, the resulting loss in Precision at Rank 5 is statistically significant according to a two-tailed T-Test (4.86\% on TDT1: p=0.5866 and 9.88\% on TREC-ROBUST: p=0.5187). Expansion terms pruning is additive to MP-RRMs. Figure 4 shows a considerable gain in efficiency over pruned RMs for every level of effectiveness. \\
\begin{figure}[h] 
\centering 
\centerline{\epsfig{file=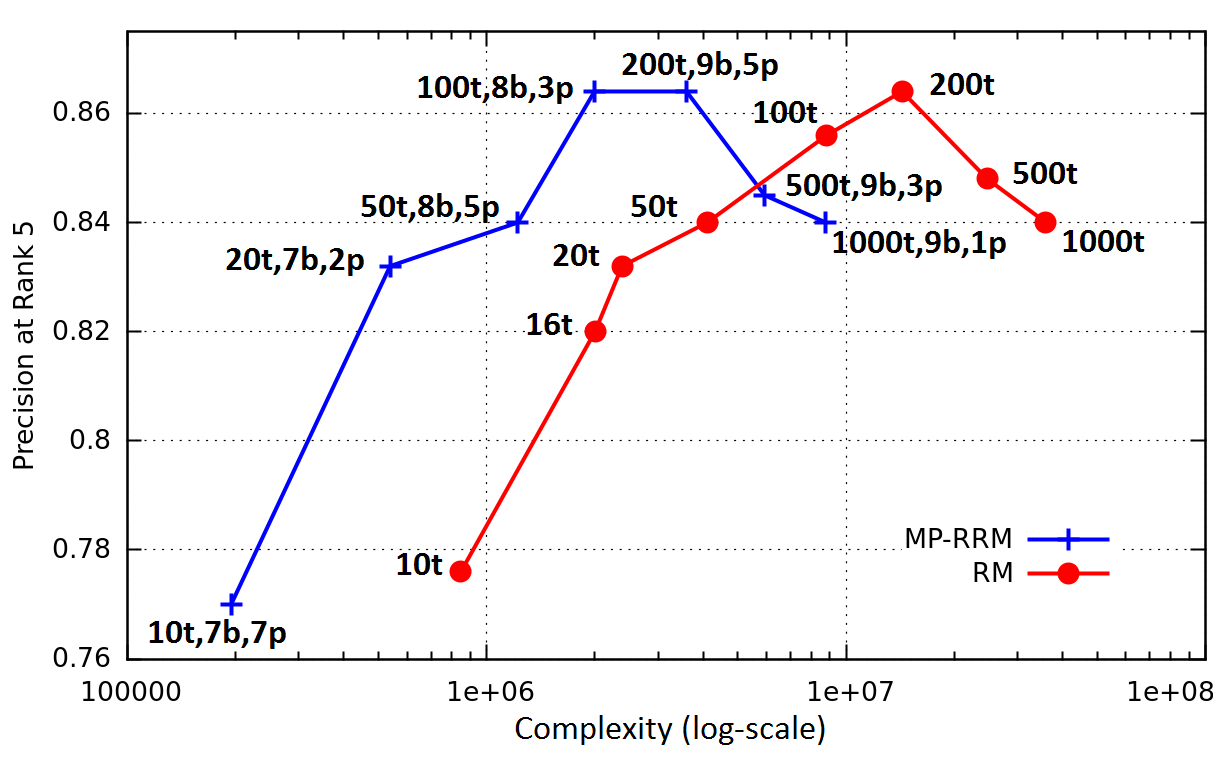, scale=0.28}} 
\vspace{-4mm} 
\caption{Effectiveness-Efficiency trade-off curve showing the impact of pruning expansion terms (t), bit-length (b) and probes (p) on 18-hash-table-MP-RRM and RM for TDT1}  
\end{figure} 
\vspace{-4mm} 
\section{Conclusion} 
We have presented an approach that tackles the inefficiency of Relevance Models for web search scenarios from the perspective of trading parametrised errors against speed increases. RRM makes use of LSH to efficiently limit the amount of documents ranked. Hence, LSH for RRM can be regarded as a different way to prune documents. While basic RRM results in modest gain, MP-RRM, which implements a variant of MPL can offer a considerable increase in speed-up over RMs. Our approach can be combined with expansion term pruning. We have shown that pruned MP-RRMs cause a reduction in complexity by up to 86.08\%, without suffering a loss in effectiveness, measured by Precision at rank 5. 
 
\end{document}